\documentclass[11pt]{article}
\usepackage{amsmath}
\usepackage{cite}
\def\be#1{\begin{equation}\label{#1}}
          \def\ee{\end{equation}}

\usepackage{pdfpages}
\usepackage[normalem]{ulem}

\title{Confining Potentials}

\author{R. L. P. G. Amaral$^{a}$\footnote{email: rubens@if.uff.br} ,
V. E. R. Lemes$^{b}$\footnote{email: vitor@dft.if.uerj.br},
O. S. Ventura$^{c}$\footnote{email: ozmrvntr@gmail.com}, L.C.Q.Vilar $^{b}$\footnote{email: lcqvilar@gmail.com}  \\
\small \em $^a$Instituto de F\'{\i}sica, Universidade Federal do Fluminense\\
\small \em Av. Litor\^anea S/N, Boa Viagem, Niter\'oi-RJ CEP. 24210-340,
Brazil\\
\small \em $^b$Instituto de F\'\i sica, Universidade do Estado do Rio de
Janeiro,\\
\small \em Rua S\~{a}o Francisco Xavier 524, Maracan\~{a}, Rio de Janeiro - RJ,
20550-013, Brazil\\
\small \em $^c$Departamento de F\'\i sica, Centro Federal de Educa\c{c}\~ao Tecnol\'ogica do Rio de
Janeiro\\
\small\em Av.Maracan\~a 249, 20271-110, Rio de Janeiro - RJ, Brazil}

\begin{document}
\maketitle
\begin{abstract}	
We use the t'Hoft-Wilson method for the generation of static fermions potential in order to
 derive  a class of confining potentials which can describe  the quark confinement.
 A general pattern for the construction of propagators through the localization of non-local actions
  is uncovered.
		\end{abstract}
	\section{Introduction}
One of the most interesting open problems of contemporary physics is
the quarks and gluons confinement \cite{weinberg}.
Confinement is a property believed to hold in QCD associated to the
non-abelian gauge symmetry, and is searched to explain the absence of colored particles from
the  QCD spectrum \cite{FSOS, FSOS2}. The role of
gauge copies in the dynamical setup underlining
the confinement has been  uncovered  years ago \cite{gribov,zwanziger,nyiri, sobreiroesorella}.

Different criteria to identify the confinement have been proposed,
according to the  distinct characteristics of the elementary particles,
such as spin and statistics.
In the gluons case, the criterion mostly used is the
breakdown of the propagator`s positivity \cite{euluizevitor1,euluizevitor2,zwanziger}, which
is a quantum mechanical requirement.
In the quarks case, confinement is associeted to the t'Hoft-Wilson criterion \cite{thoft,yndurain}.
This is ultimately quantum mechanical, althought its fundamentals can be understood classically.
It allows for the identification of the interaction potential
 and the very simple idea emerges that an infinite energy
would be necessary to separate the interacting elements.
To get a clear idea of the attractiveness of this criterion, imagine two particles at rest,
separated by the distance $ r_0 $, and submitted to the action of an
attractive Coulomb potential depending on the distance  $r$ between the charges,
\begin{equation}
U = -{1\over 4\pi}{q^2\over r}.
\label{a1}
\end{equation}
To separate these particles up to an infinite distance,
we demand the expent of an  energy minimum given by
$
\Delta U = {1\over 4\pi}{q^2\over r_0}.
\label{a2}
$
Contrast the situation, for example, with the one when the attractive potential is
\begin{equation}
U = {\alpha^2 r}.
\label{a3}
\end{equation}
In this case, an infinite amount of energy would be required to
completely separate the particles. This behavior is believed to occur in QCD
through the formation of a flux tube of color fields linking the particles. Then,
if the energy necessary to pull them apart increases indefinitely, it becomes
more energetically efficient for the system to create new particles than
to deconfine the first ones.
These are clearly confining potentials.

The static potentials are due to vectorial bosons mediated quark interactions.
This is seen through a procedure which first-quantizes
the fermions while the boson are fully second-quantized.
  As a result
  the interaction potential between two fermions of charges  $q_x$ and $q_y$
located in the spatial positions
$\vec{x}$ and $\vec{y}$
 is given by \cite{yndurain,dudal}

\begin{equation}
V(\vec r) = \lim_{T\rightarrow\infty}{{2\pi^2}\over {T}}
tr\int{d^4p}{\widetilde J^{a \mu}(p)}
{{\widetilde{M}}^{-1ab}}_{\mu\nu}(p)
{\widetilde J^{b \nu}(-p)},
\label{c4}
\end{equation}
where $\widetilde J^a_{\mu}$ is the Fourier transform of the four current
associated to the static fermions \footnote{$T^a$ are the $SU(N)$ group generators.}
\begin{equation}
	J^a_{\mu}(\vec z)=q_x\delta_{\mu 0}\delta(\vec z-\vec x)T^a+q_y\delta_{\mu 0}\delta(\vec z-\vec y)T^a
	\label{c5}
\end{equation}
 and $\vec r=\vec x-\vec y$. Here  $\widetilde M^{-1ab}_{\mu\nu}(p)$ is the inverse vector field propagator
Fourier transform. As an example, for the photon case there are no color indexes ($T^a=1$),
\begin{equation}
\widetilde M^{-1}_{\mu\nu}(p)={1\over {4\pi p^2} }\left(\delta_{\mu\nu}-
(1-\alpha){{p_\mu p_\nu}\over p^2}\right),
\label{c7}
\end{equation}
and the coulomb potential  \footnote {except for an infinite factor of self-energy}

\begin{equation}
V(\vec r) = {1\over  {4\pi}}{{q_xq_y}\over r},
\label{c16}
\end{equation}
is obtained.

The aim of this paper is to study the potentials that are associated to models that have been recently proposed.
This paper is organized as follows. In section 2, we present the propagator
that gives rise to a polynomial generalization of
the linear Cornell potential \cite{dudal,confquarks}. In section 3, we obtain the potential
 for the refined Gribov-Zwanziger propagator. In section 4, we
 review the actions associated to the Gribov-Zwanziger propagator and then obtain
 those that generate the polynomial potentials.

 \section{Polynomial Potentials}
In order to assure the confinement through the t'Hoft-Wilson criterion, the potential
must grow with the distance between the constituents. The most simple
confining potential is the linear one,
but higher polynomials can even do a better job.
Note that the dynamical input in equation (\ref{c4})
lies essentially on the propagator, which is inherent to the model.
Different propagators, however, have emerged as the result of quantum computations
of Yang-Mills theory. In this setting, interesting simple generalizations of the electromagnetic free propagator have been
considered. This prompts us to study a class of propagators that leads to polynomial potentials.

The electromagnetic-like propagator
\begin{equation}
M^{-1ab}_{\mu\nu}(p)={\delta^{ab}\over{4\pi }}\left(\frac{1}{p^{2}}+\frac{\xi m^n}{p^{n+2}}\right)\left(\delta_{\mu\nu}-(1-\alpha){{p_\mu p_\nu}\over p^2}\right)
\label{e1}
\end{equation}
will be our starting point\footnote{We introduced a constant $\xi=\pm1$. Its role will soon be clear }.
Propagators of type (\ref{e1}) appear naturally from the localization processes of actions with non local terms. The importance
 of this procedure in the study of gluon confinement has already been stressed \cite{vitor1}.
Let us include another constant $\zeta=\pm1$ to control the relative signs of the charges,  $q_x=q$ and $q_y=\zeta q$.
The current turns out to be,
\begin{equation}
\tilde J^a_{\mu}(p)={1\over {2\pi}} q \delta(p_0)\left(e^{-i\vec p.\vec x}+\zeta e^{-i\vec p.\vec {y}}\right)\delta_{\mu 0}T^a
\label{c6}
\end{equation}
and the equation (\ref{c4}) takes the form
\begin{equation}
V(\vec{x}-\vec{y}) = {{C_2(R)q^2}\over (2\pi)^3}
\int{d^3p\over p^2}\left({1}+{\xi m^n\over p^n}\right)\left(1+\zeta \cos(\vec{p}.(\vec{x}-\vec{y})\right),
\label{d2}
\end{equation}
where\footnote{$C_2(R)$ is the value of the Casimir in the
representation G.} $C_2(R)I=T^aT^a$
($I$ is the identity matrix for the
  $SU(N)$ group representation).
Keeping the notation $\vec{r}=\vec{x}-\vec{y}$ and extracting the $r$ independent self energy
piece
\begin{equation}
{q^2C_2(R)\over {2\pi^2}}\int dp\left(1+ {{\xi m^n}\over p^n}\right),
\label{d4}
\end{equation}
we obtain
\begin{equation}
V(\vec r) = {C_2(R)\zeta q^2\over {2\pi^2}}\int_{-\infty}^{\infty}
dp\left({1\over {2ipr}}{e^{ipr}}+{{\xi m^n}\over{2ip^{(n+1)}r}}e^{ipr}\right).
\label{d3}
\end{equation}
The complex integration process leads to
the potential
\begin{equation}
V(\vec r) = {\zeta C_2(R)\over  {4\pi}}{q^2\over r}+
{\xi \zeta C_2(R) i^{n}{m^{n}}\over  {4\pi}n!}{q^2 r^{n-1}}.
\label{e2}
\end{equation}
The simplest case is the Coulomb potential and it corresponds to $n=0$ or $m=0$, so that  (\ref{e1}) leads to (\ref{c7}).

The case $n=2$ is particularly important
\begin{equation}
V(\vec r) = {\zeta C_2(R)\over  {4\pi}}{q^2\over r}-{{\xi\zeta m^2}C_2(R)\over  {8\pi}}{q^2 r}.
\label{d5}
\end{equation}
This potential (Coulomb + linear) presents the standard confining properties. The case, when
$\xi=1$ and $\zeta=-1$ is known as Cornell's potential \cite{cornell}
and was obtained in \cite{dudal,confquarks}.
The presence of the linear rising term in the potential is sometimes
termed as the emergence of a "magnetic phase" of QCD,  and implies an area-law exponential
decay of the Wilson-Loop \cite{JGreensite}.
This situation is the case where the two terms, Coulomb and linear, are
attractive.

Another interesting case happens, still in the $n=2$ scenario,
when $\xi=-1$ e $\zeta=1$.
Here the Coulomb term is repulsive and the linear is attractive.
The later term is dominant for large distances
while for short distances the Coulomb one dominates.
There is an intermediate point of equilibrium which
 describes the confined system distance of stability. This gives the measure of the resultant
composed particle size.
There is allways the possibility of the existence of
other interactions, but we see that it is just enough to consider the propagator (\ref{e1})
and, of course,
just one interaction field, to obtain confinement and
stability of the composed resultant particle.

Polynomial potentials of order $n>1$, which behave at large distance as $r^n$,
lead to stronger confinement character than the linear potential.
In order to obtain the reality of the energy, $ n $ must be even.
The parity of these terms
is related to the even number of derivatives in the Lagrangians kinetic terms.

Let us stress that the existence of a balance point between the interactions,
analogous to that obtained in the $\xi=-1$ and $\zeta=1$ linear potential case,
can be obtained in a
larger set of situations. Note that in order to keep the Coulombian interaction
repulsive ($\zeta=1$),
and the polynomial part attractive,
we must impose
$\xi i^n>0$. Since $\xi=\pm1$, there is no restriction for the values of $n$, apart
from being even, which gives rise to a wide spectrum of possible forms of these confining potentials.

\section{The Refined Gribov Zwanziger Potential}
We now discuss to what extent do the
vector propagators that confine gluons by the positivity breakdown
 generate confinement of fermions.
We briefly review the potential calculation for the Gribov-Zwanziger (GZ) propagator.
For the case of pure GZ \cite{zwanziger},
\begin{equation}
\widetilde M^{-1ab}_{\mu\nu}(p)={{\delta^{ab}p^2}\over {4\pi} (p^4+\gamma^4) }\left(\delta_{\mu\nu}-(1-\alpha){{p_\mu p_\nu}\over p^2}\right),
\label{d65}
\end{equation}
the quantum corrections to the potential were analyzed in
\cite{gracey}, where $\gamma$ is determined by Gribov\textquoteright s process
and takes significance only in the infrared regime.

The  GZ model needs, however, to be refined when renormalization is taken into account.
New massive constants appear when the counter-term
is built in the quantization procedure. The propagator
takes the form \cite{gzrefinado}
\begin{align}
\widetilde M^{-1ab}_{\mu\nu}(p)={}&{\delta^{ab}(p^2+M^2)\over {4\pi} (p^4+(M^2+m^2)p^2+2g^2N\gamma^4+M^2m^2) }\nonumber\\
&  \times\left(\delta_{\mu\nu}-(1-\alpha){{p_\mu p_\nu}\over p^2}\right).
\label{d68}
\end{align}

In this case, the potential (\ref{c4}) is given by
\begin{equation}
V(\vec r) =  {C_2(R){g^2}\over{2{\pi}^2}}I,
\label{e434}
\end{equation}
where
\begin{equation}
I=\int_{-\infty}^{\infty}
{ p^2\left({p^2+M^2}\right)\over p^4+(M^2+m^2)p^2+2g^2N\gamma^4+M^2m^2 }
\left\{1+ {i\over{2pr}}e^{ipr}
\right\}dp.
\label{ewe}
\end{equation}
The self-energy results from the part
of the above integral given by
\begin{equation}
 {C_2(R){g^2}\over{2{\pi}^2}}\int_{-\infty}^{\infty}
{{p^2\left(p^2+M^2\right)}\over p^4+(M^2+m^2)p^2+2g^2N\gamma^4+M^2m^2 }
dp,
\label{ewes}
\end{equation}
and will be discarded.

If we define
\begin{equation}
\beta^2=M^2+m^2, \hspace{1,5cm} \lambda^4=2g^2N\gamma^4+M^2m^2,
\label{e43c4s7o}
\end{equation}
the potential (\ref{e434}), apart from the self-energy, takes the form

\begin{align}
V(\vec r) ={}&  {C_2(R){g^2}\over{2\pi r}}
e^{-\frac r2 \sqrt{2{\lambda^2+\beta^2}}}\left\{\frac{
2M^2-\beta^2}{2\sqrt{{4\lambda^2-\beta^2}}}\sin{\left(\frac r2\sqrt{2\lambda^2-\beta^2}\right)} +\right. \nonumber \\
{}&\left. +
\frac 12\cos{\left(\frac r2\sqrt{2\lambda^2-\beta^2}  \right)}\vphantom{\frac{
		2M^2-\beta^2}{2\sqrt{{4\lambda^2-\beta^2}}} }
\right\},
\label{e4347}
\end{align}

Taking the limit
 $\beta \rightarrow 0$ and
$M\rightarrow 0$, it results in
\begin{equation}
V(\vec r) =  {{C_2(R)g^2}\over{2\pi r}}
e^{-\lambda r {\sqrt{2} /2}}
\cos{{\lambda\sqrt{2}r}\over 2}
.
\label{e4s347}
\end{equation}
The decreasing exponential terms
present in (\ref{e4347}) and (\ref{e4s347})
points to the non confining character of the fermions in the refined Gribov-Zwanziger theory. The potential in (\ref{e4s347}) is exactly the same as presented in \cite{gracey} showing that the new scales introduced by renormalization in the refined theory do not impact the non confining character of the fermions in Gribov-Zwanziger theory.

\section{The actions}
We will provide, in this section, the actions generating the GZ
propagator and confining polynomial potentials.
The complete treatment of such actions, including the full discussion of the renormalization, will not be
dealt with here. We will restrict ourselves to show the starting point used in the
GZ renormalization, adapted to each polynomial potential generating action.

\subsection{The Gribov-Zwanziger case}
The GZ action was born  from the search for
the elimination of copies of gauge still present
 in Yang-Mills model, even after the naive fixation of the gauge, which we
assumed here to be Landau.
The starting action is\footnote{The gauge group is $SU(N)$.}
\begin{equation}
S_{\mathrm{YM}}=\int d^{4}x\,\left(\frac{1}{4}F^{\mu\nu a}F_{\mu\nu}^{a}+ib^{a}\,
\partial^{\mu}A_{\mu}^{a}+
\overline{c}^{a}\partial^{\mu}D_{\mu}^{ab}c^{b}\right)\,,\label{YM}
\end{equation}
where
\begin{equation}
F_{\mu\nu}^{a}=\partial_{\mu}A_{\nu}^{a}-\partial_{\nu}A_{\mu}^{a}+gf^{abc}A_{\mu}^{b}A_{\nu}^{c}
\end{equation}
and
\begin{equation}
D_{\mu}^{ab}=\delta^{ab}\partial_{\mu}-gf^{abc}A_{\mu}^{c}\,.
\end{equation}
The action is invariant under the BRST transformations
\begin{equation}
sA_{\mu}^{a}=-D_{\mu}^{ab}c^{b},\hspace{0.5cm}sc^{a}=
\frac{g}{2}f^{abc}c^{b}c^{c},\hspace{0.5cm}s\overline{c}^{a}=ib^{a},
\hspace{0.5cm}sb^{a}=0.\label{brs-ym}
\end{equation}
Following the Gribov procedure \cite{gribov},
the elimination of the persistent copies
is performed studying the configurations defined by the operator
\begin{equation}
\mathcal{M}^{ab}=-\partial^{\mu}D_{\mu}^{ab}\
\end{equation}
It was initially assumed that the first Gribov region ($\Omega$),
\begin{equation}
\Omega:=\{\, A_{\mu}^{a}\,|\,\partial^{\mu}A_{\mu}^{a}=0,\,
\mathcal{M}^{ab}(A)>0\,\}\,.\label{definicao}
\end{equation}
is free of copies\footnote{Now it is known that there are copies inside the
first region \cite{euluizevitor1, Baal}. The more restrictive modular region, within the first Gribov region, has to be considered in order to try to get rid of the copies.}.
As pointed out by Zwanziger \cite{zwanziger}, the implementation of the action that eliminates the copies includes a non-local term of the form
\begin{equation}
S_{\mathrm{GZ}}=S_{\mathrm{YM}}+\gamma^{4}g^{2}\int d^{4}x\, d^{4}y\,
f^{abc}A^{\mu b}(x)[\mathcal{M}^{-1}]^{ad}(x,y)f^{dec}A_{\mu}^{e}.
\end{equation}
The localization process is done by means of
the quartet
\begin{align}
 s{\bar{\omega}}_{\mu}^{a}={\bar{\varphi}}_{\mu}^{a}
\,&,& s{\bar{\varphi}}_{\mu}^{a}=0
\,,\nonumber \\
 s\varphi_{\mu}^{a}=\omega_{\mu}^{a}\,&,& s\omega_{\mu}^{a}=\,0,\label{brsgz1}
\end{align}
where $(\bar{\varphi},\varphi)$ are a pair of complex commutating
fields, while $(\bar{\omega},\omega)$ are anti-commutating ones.
The key point is the perception that
\begin{equation}
s({\bar{\omega}}^{\nu a}\partial^{\mu}D_{\mu}{{\varphi}}_{\nu}^{a})=
{\bar{\varphi}}^{\nu a}\partial^{\mu}D_{\mu}{{\varphi}}_{\nu}^{a}-
{\bar{\omega}}^{\nu a}\partial^{\mu}D_{\mu}{{\omega}}_{\nu}^{a}
+gf^{abc}\bar{\omega}_{\mu}^{a} {\partial^\nu} \left(D_{\nu}{c}^{b}{{\varphi}}^{\mu c}\right) 
\,.\label{LocalizacaoGZ}
\end{equation}
Now, the local version of the GZ action is then given by:
\begin{align}
S_{\mathrm{GZ}}^{\mbox{\footnotesize{\it local}}} &=
 S_{\mathrm{YM}}+\int d^{4}x\,\Bigl[\,-\bar{\varphi}^{\mu a}
 \partial_{\nu}D^{\nu}\varphi_{\mu}^{a}+
\bar{\omega}^{\mu a}\partial^{\nu}D_{\nu}\omega_{\mu}^{a}
+gf^{abc}\partial^\nu\bar{\omega}^{\mu a}D_{\nu}{c}^{b}{{\varphi}}_{\mu}^{c}
 \nonumber \\
&{}+{{{\gamma}^2}\over\sqrt{2}}\left(\bar{\varphi}_{\mu}^{a}+
\varphi_{\mu}^{a}\right)A^{\mu c}\Bigl]\,.\label{LocalGZ}
\end{align}
The last term of the rhs in (\ref{LocalGZ}) breaks the BRST symmetry.
To deal with this problem, Zwanziger introduced two (non dynamical)
BRST sources, also in a doublet.
These sources are taken to their physical values after
the study of renormalization  \cite{renormalizacaogz}.

A simpler way of handling the cases under consideration
emerges by taking the linear transformation of the fields
\begin{align}
 {{v}}_{\mu}^{a}&={\bar{\varphi}}_{\mu}^{a}-{{\varphi}}_{\mu}^{a}\nonumber\\
 {{u}}_{\mu}^{a}&={\bar{\varphi}}_{\mu}^{a}+{{\varphi}}_{\mu}^{a},
 \label{brsgz33}
\end{align}
so that the $v_\mu^a$ field decouples up to the quadratic terms. The action takes the form
\begin{align}
S_{\mathrm{GZ}}^{\mbox{\footnotesize{\it local}}} & =
 S_{\mathrm{YM}}+\int d^{4}x\,\Bigl[\,{-{1\over4}}{u}^{\mu a}\partial_{\nu}D^{\nu}u_{\mu}^{a}+
{1\over4}{v}^{\mu a}\partial^{\nu}D_{\nu}v_{\mu}^{a}
+\bar{\omega}^{\mu a}\partial^{\nu}D_{\nu}\omega_{\mu}^{a}+ \nonumber \\
&\phantom{=}+{g\over4}f^{abc}v^a_{\mu}\partial^{\nu}A^b_{\nu}u^{c\mu}
-{1\over2}gf^{abc}\partial^\nu\bar{\omega}^{a\mu}D_{\nu}{c}^{b}(u_{\mu}^{c}-v_{\mu}^{c}) +\nonumber \\
&\phantom{=}+{1\over\sqrt{2}}{\gamma}^2u_{\mu}^{a}A^{a\mu}\Bigl]\,\label{LocalGZ1r2}
\end{align}
and the propagators are
\begin{equation}
\langle A_{\mu}^{a}(-p)A_{\nu}^{b}(p)\rangle=
{\delta^{ab}\over {4\pi}}
{{p^2}\over{{p^4}+\gamma^4}}
\left(\delta_{\mu\nu}-\frac{p_{\mu}p_{\nu}}{p^{2}}\right),
\label{LocafghlGZ1r2}
\end{equation}
\begin{equation}
\langle A_{\mu}^{a}(-p)u_{\nu}^{b}(p)\rangle=
{\delta^{ab}\over {4\pi}}
{{m^2}\over{{p^4}+\gamma^4}}
\left(\delta_{\mu\nu}-\frac{p_{\mu}p_{\nu}}{p^{2}}\right),
\label{LocafghlGZdc1r2}
\end{equation}
\begin{equation}
\langle u_{\mu}^{a}(-p)u_{\nu}^{b}(p)\rangle=
-{\delta^{ab}\over {4\pi}}
{{p^2}\over{{p^4}+\gamma^4}}
\left(\delta_{\mu\nu}-\frac{p_{\mu}p_{\nu}}{p^{2}}\right).
\label{Locafghlfje4GZ1r2}
\end{equation}
The action (\ref{LocalGZ1r2}) presents in a more simplified way the diagonal and mixed terms,
which makes the identification of propagators more direct. Indeed, the terms that matter
 for the propagator of the vector field $A_\mu$ are the bilinear contributions $u\partial^2u$ and $uA$.
 The resource to
 $u^a_{\mu}$ and $v^a_{\mu}$, instead of the original fields  ${\bar{\varphi}}_{\mu}^{a}$ and ${{\varphi}}_{\mu}^{a}$  of the
 GZ model, simplifies the search for the actions leading to the desired propagators.

The standard procedure for the study of the renormalization of this model takes
into account the fact that (\ref{LocalGZ1r2})  can be put into the form
\begin{eqnarray}
S_{\mathrm{GZ}}^{\mbox{\footnotesize{\it local}}}  =
S_{\mathrm{YM}}+\int d^{4}x\,\Bigl[s\Omega+\Psi\Bigl]\,,
\label{LocalGxfgZ1r2}
\end{eqnarray}
where
\begin{eqnarray}
\Omega=-\frac{1}{2}\bar{\omega}^{\nu a}\partial^{\mu}D_{\mu}\left({{u}}_{\nu}^{a}-{{v}}_{\nu}^{a}\right)
\label{LocalGxfgZ1frdf2}
\end{eqnarray}
and
\begin{eqnarray}
\Psi={1\over\sqrt{2}}\gamma^2u^{\mu a}A_{\mu}^{a}.
\label{LocalGxsfgZ1r2}
\end{eqnarray}
The term (\ref{LocalGxsfgZ1r2}) is the one which breaks the
BRST symmetry. According to the standard treatment just mentioned,
BRST sources should be introduced such that
\begin{eqnarray}
sM=V \hspace{2cm} sV=0
\label{lgkjghgs}
\end{eqnarray}
and the term (\ref{LocalGxsfgZ1r2}) should be substituted by
\begin{eqnarray}
s({1\over\sqrt{2}}Mu_{\mu}^{a}A^{\mu a}).
\label{LocalGxsdfgZ1r2}
\end{eqnarray}
This means that the resultant action
is YM plus an exact  BRST variation.
The action is renormalized and, after that,  the sources are
taken to their physical values. This means,
\begin{eqnarray}
V={\gamma}^2 \hspace{2cm} M=0,
\label{lgkhhjg}
\end{eqnarray}
so that the starting action is re-obtained. In this sense, the action (\ref{LocalGZ})
should be considered as part of a larger one.

This procedure has been severely questioned recently \cite{lavrov}.
Although new improvements have been proposed defending the
 procedure \cite{silvioAh}, no definitive answer exists in the literature
 until now.
If it is valid, the GZ action can be
considered renormalized and its use is justified.
As we shall see, analogous conclusions can be reached for the case of the
actions whose potentials were
discussed in section 2.

\subsection{The linear potential action}
We take here the derivation of the GZ action presented in the previous section as a paradigm keeping the term highlighted in (\ref{LocalizacaoGZ}) but replacing the simple derivatives by covariant ones. Further, we will add the matter contribution to the transformations
(\ref{brsgz1}), resulting in
\begin{align}
  s{\bar{\omega}}_{\mu}^{a}={\bar{\varphi}}_{\mu}^{a}
+gf^{abc}c^b{\bar{\omega}}_{\mu}^{c}
\,&,\qquad s{\bar{\varphi}}_{\mu}^{a}=gf^{abc}c^b{\bar{\varphi}}_{\mu}^{c}
\,,\nonumber \\
   s\varphi_{\mu}^{a}=\omega_{\mu}^{a}+gf^{abc}c^b{{\varphi}}_{\mu}^{c}\,&,\qquad s\omega_{\mu}^{a}=
gf^{abc}c^b\omega_{\mu}^{c}\,.\label{brsgztgd1}
\end{align}
It turns out that
\begin{equation}
s({\bar{\omega}}^{\mu a}D^{\nu}D_{\nu}{{\varphi}}_{\mu}^{a})=
{\bar{\varphi}}^{\mu a}D^{\nu}D_{\nu}{{\varphi}}_{\mu}^{a}-
{\bar{\omega}}^{\mu a}D^{\nu}D_{\nu}{{\omega}}_{\mu}^{a}
\,.\label{LocalizacaoDudal}
\end{equation}
Our first task here is to construct an action that generates a propagator of the type
\begin{equation}
\langle A_{\mu}^{a}(-p)A_{\nu}^{b}(p)\rangle=
{\delta^{ab}\over {4\pi}}
\left({1\over p^2}-{{m^2}\over{p^4}}\right)
\left(\delta_{\mu\nu}-\frac{p_{\mu}p_{\nu}}{p^{2}}\right)\, ,
\end{equation}
where we have set $\xi=-1$, in (\ref{e1}),
which selects potentials
generating equilibrium points between the Coulomb and the linear term (\ref{d5}).
In terms of the fields $u^a_{\mu}$ and $v^a_{\mu}$ in (\ref{brsgz33}), we are led to consider the action:
\begin{align}
S^{\mbox{\footnotesize{\it local}}} & =
S_{\mathrm{YM}}+
\frac{1}{2}\int d^{4}x\,\Bigl[\,u^{\mu a}D^{\nu}D_{\nu}u_{\mu}^{a}-
v^{\mu a}D^{\nu}D_{\nu}v_{\mu}^{a}
\nonumber \\
&\phantom{=}-4\bar{\omega}^{\mu a}D^{\nu}D_{\nu}\omega_{\mu}^{a}+
m^2\left(v_{\mu}^{a}+A_{\mu}^{a}\right)^2\Bigl]\,.\label{Localduda2}
\end{align}
This action compares with the
 GZ case  (\ref{LocalGxfgZ1r2}) through the replacements
\begin{eqnarray}
\Omega={\bar{\omega}}^{\nu a}D^{\mu}D_{\mu}
{{(u_{\nu}^{a}-v_{\nu}^{a})}}
\label{LocalGxsfgZfd1r2}
\end{eqnarray}
and
\begin{eqnarray}
\Psi={1\over2}m^2(v_{\mu}^{a}+A_{\mu}^{a})^2.
\label{LocalGxsfgZd1r2}
\end{eqnarray}
The fundamental aspect of the renormalizability of this model
is now being investigated,
as well as the GZ case, and will be left for further discussion.

Let us stress a point that bears some similarity to the GZ case.
Considering the use of the equations of motion
for the auxiliary fields, we see that (\ref{Localduda2})
originates from the localization process of a
non-local action of the form
\begin{equation}
S^{\widetilde n\mbox{\footnotesize{\it local}}}  =
S_{\mathrm{YM}}+
{1\over2}m^2 \int d^{4}xA^{\mu a} {\cal N}\delta^{ab}\delta_{\mu\nu}
A^b_{\nu}
\,,\label{naoLocaldudal1}
\end{equation}
where
\begin{equation}
{\cal N } =1+ {m^2\over{D^2}}
-{m^4\over{4(D^2+m^2)^2}}+{m^6\over{4D^4(D^2+m^2)}}
	\,.\label{naoLocaldudal2}
	\end{equation}

Observe the similarity of this linear potential case  with the GZ case in which two
doublets were also needed, $\{\bar{\varphi}_{\mu}^{a},\bar{\omega}_{\mu}^{a}\}$ and
$\{{\varphi}_{\mu}^{a}, \omega_{\mu}^{a}\}$, as shown in
 (\ref{LocalGZ}).  The structure of treatment is the same. We will see that for non-linear potentials cases the situation is quite different

\subsection{Cubic potential action}

We will discuss how to obtain the action for the more confining cubic potential
\begin{equation}
V(\vec r) = {C_2(R)\zeta\over  {4\pi}}\left(
{q^2\over r}+
{\xi{m^{4}}\over  4!}{q^2 r^{3}}\right),
\label{e1d2}
\end{equation}
that is generated by the gauge field propagator of the form
\begin{equation}
\langle A_{\mu}^{a}(-p)A_{\nu}^{b}(p)\rangle=
{\delta^{ab}\over {4\pi}}
\left({1\over p^2}+{\xi{m^4}\over{p^6}}\right)
\left(\delta_{\mu\nu}-\frac{p_{\mu}p_{\nu}}{p^{2}}\right)\, ,
\end{equation}
which can be obtained from (\ref{e2}) and (\ref{e1}) respectively by chosing $n=4$.
First, we consider the case where $\xi = 1$.
Within the same spirit of the last section we consider now the action
\begin{align}
S^{\mbox{\footnotesize{\it local}}} & =
S_{\mathrm{YM}}+
\frac{1}{2}\int d^{4}x\,\Bigl[
v^{\mu a}D^{\nu}D_{\nu}v_{\mu}^{a}-
u^{\mu a}D^{\nu}D_{\nu}u_{\mu}^{a}
+4\bar{\omega}^{\mu a}D^{\nu}D_{\nu}\omega_{\mu}^{a}
\nonumber \\
&\phantom{=\frac{1}{2}\int }+  \,m^2\left(A_{\mu}^{a}+u_{\mu}^{a}\right)v^{\mu a}
\Bigl].
\label{lgkjg1}
\end{align}

The terms with double covariant derivatives of the auxiliary fields in (\ref{lgkjg1}) are
very similar to those in the linear potential case.
They
are seen to be BRST trivial by the use of (\ref{LocalizacaoDudal}).
In this sense the possibility of the BRST quantization should take into account
the replacement of (\ref{LocalGxfgZ1frdf2}) by
\begin{eqnarray}
\Omega={-\bar{\omega}}^{\nu a}D^{\mu}D_{\mu}{{(u_{\nu}^{a}-v_{\nu}^{a})}}
\label{LocalGxsfgZfd1r2}
\end{eqnarray}
and (\ref{LocalGxsfgZ1r2}) by
\begin{eqnarray}
\Psi={1\over2}m^2v^{\mu a}(A_{\mu}^{a} +u_{\mu}^{a} ).
\label{LocalGxfsfgZd1r2}
\end{eqnarray}
Again, the last term of (\ref{lgkjg1}) breaks the
BRST symmetry.

The genesis of the action (\ref{lgkjg1}) from the localization of a non-local action,
analogously to the linear case (\ref{naoLocaldudal1}), implies  ${\cal N}$ to be
  \begin{eqnarray}
{\cal N}  = {m^6D^2\over{(D^4+{m^4})^2}}
		\,.\label{naoLocalddudal2}
	\end{eqnarray}

Let's look now at the case $\xi = -1$. This case is more interesting because it provides a point
of equilibrium between the contributions of the two terms of the potential, i.e., the force coming from
the Coulombian term cancels the originated from the cubic term, what can lead to a prediction of the nucleon radius.

In this case the kinetic terms of the localizing fields  have no relative sign.
This implies the impossibility in constructing the kinetic term of the auxiliary fields sector
with a quartet. Indeed we need two quartets of localizing fields. They
are  introduced with the transformations below,

\begin{align}
   s{\bar{\omega}}_{\mu}^{a}&={\bar{\varphi}}_{\mu}^{a}
+gf^{abc}c^b{\bar{\omega}}_{\mu}^{c}
\,,\qquad s{\bar{\varphi}}_{\mu}^{a}=gf^{abc}c^b{\bar{\varphi}}_{\mu}^{c}
\,,\nonumber \\
   s\varphi_{\mu}^{a}&=\omega_{\mu}^{a}+gf^{abc}c^b{{\varphi}}_{\mu}^{c}\,,\qquad s\omega_{\mu}^{a}=gf^{abc}c^b\omega_{\mu}^{c}\,,\nonumber\\
   s{\bar{\xi}}_{\mu}^{a}&={\bar{\psi}}_{\mu}^{a}
+gf^{abc}c^b{\bar{\xi}}_{\mu}^{c}
\,,\qquad s{\bar{\psi}}_{\mu}^{a}=gf^{abc}c^b{\bar{\psi}}_{\mu}^{c}
\,,\nonumber \\
   s\psi_{\mu}^{a}&=\xi_{\mu}^{a}+gf^{abc}c^b{\psi}_{\mu}^{c}\,,\qquad s\xi_{\mu}^{a}=gf^{abc}c^b\xi_{\mu}^{c}\,.\label{brsgz2ed1s}
\end{align}
Note that
\begin{eqnarray}
&&s({\bar{\omega}}^{\nu a}D^{\mu}D_{\mu}{{\varphi}}_{\nu}^{a})=
{\bar{\varphi}}^{\nu a}D^{\mu}D_{\mu}{{\varphi}}_{\nu}^{a}-
{\bar{\omega}}^{\nu a}D^{\mu}D_{\mu}{{\omega}}_{\nu}^{a}\\
&&s({\bar{\xi}}^{\nu a}D^{\mu}D_{\mu}{{\psi}}_{\nu}^{a})=
{\bar{\psi}}^{\nu a}D^{\mu}D_{\mu}{{\psi}}_{\nu}^{a}-
{\bar{\xi}}^{\nu a}D^{\mu}D_{\mu}{{\xi}}_{\nu}^{a}
\,.\label{LocalizacaoGZf}
\end{eqnarray}
Finally, introducing the field transformations
\begin{align}
{{v}}_{\mu}^{a}&={\bar{\varphi}}_{\mu}^{a}-{{\varphi}}_{\mu}^{a}\qquad
{{u}}_{\mu}^{a}={\bar{\varphi}}_{\mu}^{a}+{{\varphi}}_{\mu}^{a}.\nonumber\\
{{k}}_{\mu}^{a}&={\bar{\psi}}_{\mu}^{a}-{{\psi}}_{\mu}^{a}\qquad
{{t}}_{\mu}^{a}={\bar{\psi}}_{\mu}^{a}+{{\psi}}_{\mu}^{a},
\label{brsgz33fgbh}
\end{align}
the action we are looking for takes the form
\begin{align}
S^{\mbox{\footnotesize{\it local}}} & =
S_{\mathrm{YM}}+
\frac{1}{2}\int d^{4}x\,\Bigl[
v^{\mu a}D^{\nu}D_{\nu}v_{\mu}^{a}-u^{\mu a}D^{\nu}D_{\nu}u_{\mu}^{a}+
k^{\mu a}D^{\nu}D_{\nu}k_{\mu}^{a}-t^{\mu a}D^{\nu}D_{\nu}t_{\mu}^{a}+\nonumber\\
&\phantom{=}
- 2\bar{\omega}^{\mu a}D^{\nu}D_{\nu}\omega_{\mu}^{a}+
-2\bar{\xi}^{\mu a}D^{\nu}D_{\nu}\xi_{\mu}^{a}+
+ 2m^2A^{\mu a}v_{\mu}^{a}+ 2m^2k^{\mu a}v_{\mu}^{a}
\Bigl].
\label{lgkjg3d4}
\end{align}
This action localizes a nonlocal term which,
in the form (\ref{naoLocaldudal1}), has ${\cal N} $ given by
\begin{equation}
{\cal N}  =\frac 14(D^4+m^4)\left({mD\over{2D^4-m^4}}\right)^2
\,.\nonumber\label{naoLocaldudadfsgl2}
\end{equation}
The similarity with the GZ case can be seen if we look at
(\ref{LocalGxfgZ1r2}) taking
\begin{eqnarray}
\Omega={-2\bar{\omega}}^{\nu a}D^{\mu}D_{\mu}{{\varphi}}_{\nu}^{a}
-2\bar{\xi}^{\nu a}D^{\mu}D_{\mu}{{\psi}}_{\nu}^{a}
\label{LocalGxsfgZffd1r2}
\end{eqnarray}
and
\begin{eqnarray}
\Psi=2A^{\mu a}v_{\mu}^{a}+ 2k^{\mu a}v_{\mu}^{a}.
\label{LocalGxsfgkhgZfsdd1r2}
\end{eqnarray}

The generalization to actions generating arbitrary even powers in $1/p^{n+2}$ written in (\ref{e1}) (with $\xi=-1$), demands $n$ doublets. In the case presently analyzed we
needed four doublets and just two of these fields couple to the gauge field.

\section{Conclusion}
We investigate in this work the t'Hoft - Wilson method \cite{thoft}
and its application to the construction of potentials
arising from interactions  between fermions, mediated
by gauge fields.
With the use of the first and second quantization processes,
one is able to identify the potential of static charges in general cases.
We apply these ideas to calculate the potential for the refined GZ model and to generalize the linear Cornell potential
 \cite{cornell} to polynomial ones $r^{n-1}$,
being $n$ an even number.

Within this scenario we establish the interaction described
by a vector field $A_{\mu}$ that
can lead to situations of quarks confinement.
In general these contributions are non local terms added to the Yang-Mills action.

In particular, along our analysis, we showed the non-confining character of the refined GZ model in the sense of the 't Hooft-Wilson criteria.
Anyway, inspired by the GZ action, we used the same localization process to build local actions leading to the polinomial confining potentials that we studied previously.
The quartet structure that is typical of the GZ construction then needs to be adapted and generalized in order to generate confining potentials taking into account the necessity of a point of stable equilibrium.
This will be associated to the average radius of the fermionic condensed state.

Obviously, when we look at the legitimacy of such Lagrangians, other physical demands need to be considered.
The renormalizability of these models becomes a fundamental question.
An important point is that the Zwanziger-Sorella treatment of the renormalization that we briefly showed in section (4.1) is now being questioned by recent results  \cite{lavrov}.
Although an alternative treatment has been constructed \cite{silvioAh} (with the introduction of a Stuckelberg auxiliary field) we think that  the renormalization of these models is still an open question.

It is worth mentioning that the calculation of the potential for the original GZ model was done in \cite{gracey} and showed its nonconfining character.
As we said a moment ago the improved GZ theory displays the same feature. We understand that
this happens because the propagator that breaks the positivity, a basic ingredient for gluon confinement, does not generate
a potential which increases with the distance between the fermions.
But our ultimate goal was acchieved,  as we succeded  in describing local actions with a confining character in 't Hooft sense by
the use of the same machinery characteristic of the GZ construction.
This opens a possible path in the direction of building an action with both effects in the same context, aiming a simultaneous  confinement of quarks and gluons.

\end{document}